\documentclass[11pt]{article}

\usepackage[margin=1in]{geometry}
\usepackage{amsfonts, amsmath, amssymb}
\usepackage{array, graphicx}
\usepackage[numbers]{natbib}
\usepackage{caption}
\usepackage{subcaption}
\usepackage{booktabs}
\usepackage[utf8]{inputenc}
\usepackage{graphicx}

\title{An analysis of network filtering methods to sovereign bond yields during COVID-19}

\author{Raymond Ka-Kay Pang
\thanks{London School of Economics and Political Science, Department of Mathematics, Houghton Street, London WC2A 2AE, UK, Email: r.pang@lse.ac.uk. }
\and 
Oscar Granados
\thanks{Department of Economics and International Trade, Universidad Jorge Tadeo Lozano, Bogot\'a, Colombia, Email: oscarm.granadose@utadeo.edu.co}
\and
Harsh Chhajer
\thanks{Centre for Biosystems Science and Engineering, Indian Institute of Science, Bangalore 560012, India, Email: harshchhajer@iisc.ac.in}
\and
Erika Fille Legara
\thanks{Aboitiz School of Innovation, Technology, and Entrepreneurship, Asian Institute of Management, Makati, Philippines, Email: elegara@aim.edu}
}

\begin{document}
\maketitle

\begin{abstract}
In this work, we investigate the impact of the COVID-19 pandemic on sovereign bond yields. We consider the temporal changes from financial correlations using network filtering methods. These methods consider a subset of links within the correlation matrix, which gives rise to a network structure. We use sovereign bond yield data from 17 European countries between the 2010 and 2020 period. We find the mean correlation to decrease across all filtering methods during the COVID-19 period. We also observe a distinctive trend between filtering methods under multiple network centrality measures. We then relate the significance of economic and health variables towards filtered networks within the COVID-19 period. Under an exponential random graph model, we are able to identify key relations between economic groups across different filtering methods.  

\end{abstract}


\begin{center}
\end{center}

\maketitle


\section{Introduction}
The novel coronavirus disease 2019 (COVID-19) epidemic caused by SARS-CoV-2 began in China in December 2019 and rapidly spread around the world. The confirmed cases increased in different cities of China, Japan, and South Korea in a few days of early January 2020, but spread globally with new cases in Iran, Spain, and Italy within the middle of February. We focus on sovereign bonds during the COVID-19 period to highlight the extent to which the pandemic has influenced the financial markets. A sovereign bond is a bond that is issued by sovereign entities or administrative regions. The yield of these bonds is the interest rate which is paid to the buyer of the bond by the issuer. Each issued sovereign bond has an associated maturity date and is considered risk-free. However, the yields of sovereign bonds can depend highly on factors such as the inflation, political stability, and the debt of the issuing country.\

In the last few years, bond yields across the Euro-zone were decreasing under a range of European Central Bank (ECB) interventions, and overall remained stable compared with the German Bund, a benchmark used for European sovereign bonds. These movements were disrupted during the COVID-19 pandemic, which has affected the future trajectory of bond yields from highly impacted countries, e.g., Spain and Italy. However, in the last months, the European central banks intervened in financial and monetary markets to consolidate stability through an adequate supply of liquidity countering the possible margin calls and the risks of different markets and payment systems. These interventions played a specific role in sovereign bonds because, on the one side, supported the stability of financial markets and, on the other side, supported the governments' financial stability and developed a global reference interest rate scheme. Understanding how correlations now differ and similarities observed in previous financial events are important in dealing with the future economic effects of COVID-19.\

We consider an analysis of sovereign bonds by using network filtering methods, which is part of a growing literature within the area of econophysics~\cite{Li2019, Stavroglou2016, maeng2012, leon2014, gilmore2010}. The advantage in using filtering methods is the extraction of a network type structure from the financial correlations between sovereign bonds. Hence, the correlation-based networks and hierarchical clustering methodologies allow us to understand the nature of financial markets and some sovereign bond features. It is not clear which approach should be used in analyzing sovereign bond yields, and so within this paper, we implement various filtering methods to the sovereign bond yield data and compare the resulting structure of different networks. Through this analysis, we able to evaluate the impact which the topological structure of filtered networks has on the economic and health relations between nodes.\

Our results show that the mean correlation peaks in October 2019 and then decreases during the 2020 period, when COVID-19 is most active in Europe. These dynamics are reflected across all network filtering methods and represent the wide impact of COVID-19 on the spectrum of correlations, compared to previous financial events. We also find a clustering of Euro-area countries and a disintegration with non-Euro countries during the COVID-19 period. These network structures reflect the financial state of sovereign bonds observed within previous financial events but are also related by exogenous variables, e.g., death rates of countries, which we can analyze under an exponential random graph model.\

Previous studies have used different methods to analyze historic correlations as random matrix theory to identify the distribution of eigenvalues concerning financial correlations~\cite{laloux2000random,plerou2002random,junior2012correlation}, the partial transfer entropy to quantify the indirect influence that stock indices have on one another~\cite{Junior_2015}, the approaches from information theory in exploring the uncertainty within the financial system ~\cite{huang2012multifractal,darbellay2000entropy}, community structure analysis~\cite{Vodenska_2016}, multilayer network methods~\cite{ALDASORO201817,Bargigli2016, TONZER201519,RePEc:ecb:ecbwps:20161944,GULEVA2015257, POLEDNA201570}, and filtering methods. 
Several authors have used network filtering methods to explain financial structures~\citep{mantegna1999hierarchical,onnela2003asset}, hierarchy and networks in financial markets \citep{TUMMINELLO201040}, relations between financial markets and real economy \citep{10.1371/journal.pone.0116201}, volatility \citep{doi:10.1080/14697688.2018.1535183}, interest rates
\citep{DIMATTEO200521}, stock markets \citep{Isogai2017,Wang2017,Wu2018,10.1007/978-3-319-99010-1_53}, future markets \citep{Bartolozzi2007} or topological dynamics \citep{Tang2018} to list a few. Also, the comparison of filtering methods to market data has been used for financial instruments. Birch, et al~\cite{birch2016analysis} consider a comparison of filtering methods of the DAX30 stocks. Musmeci, et al~\cite{musmeci2017multiplex} propose a multiplex visual network approach and consider data of multiple stock indexes. Kukreti, et al~\cite{kukreti2020perspective} use the S\&P500 market data and incorporate entropy measures with a range of network filtering methods. Aste, et al~\cite{aste2010correlation} apply a comparison of network filtering methods on the US equity market data and assess the dynamics using network measures, Schwendner, et al~\cite{Schwendneretal2015} applied a correlation influence approach and constructed noise-filtered influence networks to understand the collective yield dynamics of the Euro area sovereign bonds.\

To evaluate the European sovereign bonds based on filtering methods, this work is organized as follows. In Section \ref{sec:meth}, we describe the network filtering methods and present the data sets with some preliminary empirical analyses. We apply in Section \ref{sec: results} the filtering methods to sovereign bond yields and analyze the trend of financial correlations over the last decade, and consider aspects of the network topology. We construct plots in Section \ref{sec: plots} representing the COVID-19 period and consider an analysis using the exponential random graph model for each filtering method. In Section \ref{sec: con}, we discuss the results and future directions.\

\section{Materials and methods}
\label{sec:meth}

We introduce a range of network filtering methods and consider a framework as in \cite{mantegna1999hierarchical} for sovereign bond yields. We define $n\in\mathbb{N}$ to be the number of sovereign bonds and bond yields $Y_{i}(t)$ of the $i$th sovereign bond at time-t, where $i\in\{1,...,n\}$. The correlation coefficients $r_{ij}(t)\in[-1,1]$ are defined using Pearson correlation as\  

\begin{equation}
r_{ij}=\frac{\langle Y_{i}Y_{j}\rangle - \langle Y_{i}\rangle\langle Y_{j}\rangle}{\sqrt{\left(\langle Y^{2}_{i}\rangle-\langle Y_{i} \rangle^2\right)\left(\langle Y^{2}_{j}\rangle-\langle Y_{j} \rangle^2\right)}},
\end{equation}\

with $\langle\cdot\rangle$ denoting the average of yield values. The classical approach in using the Pearson correlation is well established, but it does not take into account the increases in correlation from market volatility. We can account for these changes by considering the conditional Pearson correlation approach as in \cite{forbes2002no}. We define an adjustment factor $\beta_{ij}(t)\in[0,\infty)$ and the conditional correlation $r^*_{ij}$ at time-t as follows:\

\begin{equation}
r^*_{ij}=r_{ij}\sqrt{\frac{1+\beta_{ij}}{1+\beta_{ij}r_{ij}^2}},\ \textrm{where}\ \beta_{ij}=\frac{\sigma^{h}_{ij}}{\sigma^{l}_{ij}}-1.
\end{equation}\

This adjustment factor is represented by the relative difference between two subgroups of high covariance $\sigma^h_{ij}(t)\in[0,\infty)$ and low covariance $\sigma^l_{ij}(t)\in[0,\infty)$ of bond yields at time-t. As the relative difference in covariance increases, this increases the adjustment factor $\beta_{ij}(t)$ and the magnitude of the conditional Pearson correlation. This adjustment preserves the symmetry of correlation values between sovereign bonds $i$ and $j$, while taking into account market conditions. We form both subgroups by equally dividing yield values, where the high variance $\sigma^{h}_{ij}$ group consists of the $25\%$ lowest and highest yield values, with the remaining values allocated within the low variance $\sigma^{l}_{ij}$ group. This allocation is applied individually to each sovereign bond in which the covariance is computed.\

Under the conditional Pearson correlation, we establish the notion of distance $d_{ij}\in[0,2]$. We consider the values of the entries $r^*_{ij}$ on the conditional correlation matrix $R^*\in [-1,1]^{n\times n}$, with $d_{ij}=\sqrt{2(1-r^*_{ij})}$. A distance of $d_{ij}=0$ represents perfectly positive correlations and $d_{ij}=2$ represents bonds with negative correlations. The network filtering methods are then applied to the distance matrix $D\in [0,2]^{n\times n}$, where a subset of links (or edges) are chosen under each filtering method. The set of edges is indicated by $\{(i,j)\in E(t):$ nodes $i$ and $j$ are connected$\}$ at time-t, defined for each filtering method. We define the time frames of financial correlations as $X$ for the set of observations, with $n$ different columns and $T$ rows. From the set of observations $X$, we consider windows of length $120$, which is equal to six months of data values. We then displace $\delta$ windows by $10$ data points, which is equal to two weeks of data values, and discard previous observations until all data points are used. By displacing the data in this way, we can examine a time series trend between each window $X$.\

\subsection{Network filtering methods}

We consider multiple network's filtering methods to analyze the dynamics from multiple perspectives. We introduce the commonly used minimum spanning tree (MST) method, which has been used within currency markets \cite{jang2011currency}, stocks markets \cite{sandoval2012pruning,situngkir2005stock} and sovereign bond yields \cite{dias2012sovereign}. The MST from Table \ref{tab:listfilt} considers the smallest edges and prioritizes connections of high correlation to form a connected and undirected tree network. These networks can be constructed from a greedy type algorithm e.g. Kruskal's and Prim's algorithm and satisfies the properties of subdominant ultrametric distance e.g. $d_{ij}\leq\max\{d_{ik},d_{kj}\}\ \forall i,j,k\in\{1,...,n\}$.\ 

This approach is used as it establishes three key properties within a subset of correlations. We argue these properties are relevant within filtering methods but can also be individually constrained when applied in conjunction (as within the MST) for topological and economic reasons. By considering four methods, we can analyze the influence of each feature on the properties of the network: 

\begin{itemize}
    \item \textbf{Connectivity:} Under the MST, all nodes are connected within the network. As there has been a broad impact from COVID-19, many sovereign bonds have experienced a comovement in yield trends under market conditions. This criterion in which the network structure is connected also excludes some highly positive links and decreases the information between positively correlated sovereign bonds. Therefore, we consider the Asset Graph (AG), which includes all positive correlations of interest while maintaining the network density.
    \item \textbf{Sparsity:} The key motivation in filtering methods is the decrease in links, in which we can establish network properties of interest, e.g., network centrality. As observed in the 2012 Euro debt crisis, specific sovereign bonds are large contributors to the spillover effects observed in other bond yield trends. The fixed number of links within the MST can be also argued to oversimplify the network and reduce connectivity. Hence, we consider the Triangulated Maximal Filtering Graph (TMFG), which establishes a planar graph and increases the total number of links compared with the MST.   
    \item \textbf{Positivity:} From an economic perspective, positive correlations are relevant in identifying the trends in different financial instruments and periods of high volatility. However, focusing on this subset of correlations may exclude sovereign bonds that act differently, i.e., although the majority of bond yields increase within the COVID-19 period, several bond yields like Germany and Switzerland decrease. To account for these dynamics, we consider a Maximum Spanning Tree (MaST), which prioritizes negative correlations within the network.    
\end{itemize}

\begin{table}[ht]
\centering
\begin{tabular}{|>{\centering\arraybackslash}m{3.5cm}|>{\centering\arraybackslash}m{2.5cm}|>{\centering\arraybackslash}m{2cm}|>{\arraybackslash}m{6cm}|} 
\hline
\textbf{Network Filtering Methods} & \textbf{Number of links (edges)} & \textbf{Reference} & \textbf{Description}\\
\hline
Minimum Spanning Tree (MST) & $n-1$ & \cite{kruskal1956shortest} & A connected and undirected network for $n$ nodes which minimizes the total edge weight.\\
\hline
Maximum Spanning Tree (MaST) & $n-1$ & \cite{qian2010universal} & A connected and undirected network for $n$ nodes which maximizes the total edge weight.\\
\hline
Asset Graph (AG) & $n-1$ & \cite{onnela2003dynamic} & Choose the smallest $n-1$ edges from the distance matrix.\\
\hline
Triangulated Maximal Filtering Graph (TMFG) & $3(n-2)$ &\cite{massara2016network} & A planar filtered graph under an assigned objective function.\\
\hline
\end{tabular}
\caption{List of network filtering methods.}
\label{tab:listfilt}
\end{table}

We provide further descriptions of the methods described above. An AG considers positive correlations between nodes of a given threshold. All $n-1$ highest correlations are considered in an AG, allowing for the formation of cliques not observed within a MST network. The use of AG has been considered in Onnela, et al~\cite{onnela2004clustering}, which identifies clustering within stock market data. As the method only considers $n-1$ links, some nodes within the AG may not be connected for the given threshold. Therefore, the connection of unconnected nodes is unknown, relative to connected components.\ 

The TMFG constructs a network of $3(n-2)$ fixed edges for $n$ nodes, similar to the planar maximal filtered graph (PMFG) \cite{tumminello2005tool}, which has been used to analyze US stock trends \cite{musmeci2017multiplex}. The algorithm initially chooses a clique of 4 nodes, where edges are then added sequentially, in order to optimize the objective function e.g., the total edge weight of the network, until all nodes are connected. This approach is non-greedy in choosing edges and incorporates the formation of cliques within the network structure. A TMFG is also an approximate solution to the weighted planar maximal graph problem, and is computationally faster than the PMFG. The resulting network includes more information about the correlation matrix compared with spanning tree approaches.\

The MaST constructs a connected and undirected tree network with $n-1$ edges in maximizing the total edge weight. Analyses involving MaST have been used as comparisons to results observed within MST approaches \cite{dias2013spanning,heimo2009maximal}. An MaST approach is informative for connections of perfectly anti-correlation between nodes, which are not displayed within the MST.\

\subsection{Sovereign bond data}

The European sovereign debt has evolved in the last ten years, with some situations affecting the convergence between bond yields. After the 2008 crisis, European countries experienced a financial stress situation starting in 2010 that affected bond yields. Thus, the investors saw an excessive amount of sovereign debt and demanded higher interest rates in low economic growth situations and high fiscal deficit levels. During 2010-2012, several European countries suffered downgrades in their bond ratings to junk status that affected investors' trust and fears of sovereign risk contagion resulting, in some cases, a differential of over 1,000 basis points in several sovereign bonds. After the introduction of austerity measures in GIIPS (Greece, Ireland, Italy, Portugal, and Spain) countries, the bond markets returned to normality in 2015.\

The 2012 European debt crisis revealed spillover effects between different sovereign bonds, which have been studied using various time series models, e.g., VAR \cite{claeys2014measuring,antonakakis2013sovereign} and GARCH \cite{balli2009spillover}. The results showed that Portugal, Greece, and Ireland have a greater domestic effect, with Italy and Spain contributing to the spillover effects in other European bond markets. A core group of ABFN (Austria, Belgium, France, and the Netherlands) countries had a lower contribution to the spillover effects, with some of the least impacted countries residing outside of the Eurozone.\

\begin{table}[ht]
\centering
\begin{tabular}{ccccccccccc}
  \hline
 \multicolumn{1}{c}{\textbf{Country}} &
  \multicolumn{1}{c}{\textbf{Min}} &
   \multicolumn{1}{c}{\textbf{Max}} &
 \multicolumn{1}{c}{\textbf{Mean}} & \multicolumn{1}{c}{\textbf{Variance}}&
\multicolumn{1}{c}{\textbf{Skewness}} & \multicolumn{1}{c}{\textbf{Kurtosis}}& 
\multicolumn{1}{c}{\textbf{AC(1)}} & \multicolumn{1}{c}{\textbf{AC$^2$(1)}}\\
  \hline
Austria & -0.47 & 3.90 & 1.30 & 1.45 & 0.55 & 2.11 & 0.05 & 0.14 \\ 
  Belgium & -0.43 & 5.83 & 1.58 & 2.01 & 0.61 & 2.13 & 0.17 & 0.41 \\ 
  Czech & 0.24 & 4.55 & 1.88 & 1.32 & 0.61 & 2.33 & -0.07 & 0.11 \\ 
  France & -0.44 & 3.79 & 1.40 & 1.37 & 0.40 & 1.91 & 0.06 & 0.18 \\ 
  Germany & -0.85 & 3.50 & 0.95 & 1.13 & 0.57 & 2.42 & -0.01 & 0.21 \\ 
  Greece & 0.56 & 39.85 & 9.16 & 51.02 & 1.74 & 6.31 & 0.07 & 0.01 \\ 
  Hungary & 1.55 & 10.73 & 4.67 & 4.62 & 0.55 & 1.90 & 0.01 & 0.22 \\ 
  Iceland & 2.19 & 8.15 & 5.72 & 1.73 & -0.88 & 3.14 & 0.04 & 0.17 \\ 
  Ireland & -0.32 & 14.45 & 2.82 & 9.17 & 1.23 & 3.65 & -0.36 & 0.50 \\ 
  Italy & 0.48 & 7.31 & 2.96 & 2.40 & 0.55 & 2.28 & 0.07 & 0.08 \\ 
  Netherlands & -0.64 & 3.78 & 1.16 & 1.28 & 0.50 & 2.16 & 0.01 & 0.20 \\ 
  Poland & 1.15 & 6.40 & 3.68 & 1.87 & 0.34 & 2.23 & 0.05 & 0.16 \\ 
  Portugal & -0.05 & 17.36 & 4.30 & 11.94 & 1.10 & 3.54 & -0.28 & 0.32 \\ 
  Romania & 2.56 & 10.80 & 4.91 & 2.20 & 0.73 & 2.60 & -0.37 & 0.30 \\ 
  Spain & -0.01 & 7.56 & 2.70 & 3.60 & 0.52 & 1.91 & 0.13 & 0.14 \\ 
  Switzerland & -1.11 & 2.14 & 0.33 & 0.57 & 0.64 & 2.42 & 0.02 & 0.08 \\ 
  UK & 0.07 & 4.28 & 1.81 & 0.91 & 0.42 & 2.61 & -0.04 & 0.17 \\ 
   \hline
\end{tabular}
\caption{Summary statistics of the 10Y sovereign bond yield data of 17 European countries from January 2010 to December 2020. AC(1) represents the first-order autocorrelation of the difference between yield values and AC$^2$(1) represents the first-order autocorrelation of the squared series.}
\label{tab:StatTab}
\end{table}

During the sovereign debt crisis, public indebtedness increased after Greece had to correct the public finance falsified data, and other countries created schemes to solve their public finance problems, especially, bank bailouts. In consequence, the average debt-to-GDP ratio across the Euro-zone countries rose from 72\% in 2006 to 119.5\% in 2014, as well as the increase in sovereign credit risk \citep{ALTER2014134,BECK2016449}.\ 

After the Fiscal Compact Treaty went into effect at the start of 2013, the yield of sovereign bonds started a correction. This treaty defined that fiscal principles had to be embedded in the national legislation of each country that signed the treaty. Although some investors and institutions pushed for financial and monetary authorities to introduce an additional decision, that permitted them to include sovereign bonds in their portfolios. The rate interest policy of the European Central Bank helped to consolidate the trust in this kind of asset; the bonds confirmed their adjustment especially Germany, France, Spain, during the fourth quarter of 2013, while countries like Greece and Italy started in 2014 with variations of over 500 basis points during the following months. By 2015, all European bonds increased their yields as a result of an adjustment of the market rally of 2014.\

We analyze the sovereign bond yield data for the following countries Austria (AUT), Belgium (BEL), Czech Republic (CZE), France (FRA), Germany (DEU), Greece (GRC), Hungary (HUN), Iceland (ISL), Ireland (IRL), Italy (ITA), Netherlands (NLD), Poland (POL), Portugal (PRT), Romania (ROU), Spain (ESP), Switzerland (CHE), and the UK (GBR). We consider sovereign bond yields with a 10 year maturity between January 2010 and Dec 2020. This data is taken from the financial news platform Investing \cite{investing}. In total, there are 2,615 data values for each country with an average of 238 data points within 1 year.\

Table \ref{tab:StatTab} provides summary statistics of the 10Y bond yield data. The data shows that the lowest recorded yields for many countries was within 2020, during the COVID-19 period and highest in 2011, before the 2012 European debt crisis. The lowest yield values are with Germany and Switzerland, which both record yield values lower than $-0.80$. In contrast, Greece has the highest yield value of $39.85$ and a variance of $51.02$. The left skewed yield distributions (except for Iceland) represent an average decrease in yield values and are high for GIIPS countries compared with the UK, France, and Germany, with flattening yield trends. If we examine the autocorrelation, we find this to be small overall but high for some countries e.g., Belgium, Ireland and Portugal within the squared series.\

\section{Network measures}
\label{sec: results}

We compute the correlation matrix for each window $X$ with a displacement of $\delta$ between windows, and consider the mean and variance for the correlation matrix. We define the mean correlation $\overline{r}(t)$ given the conditional correlation values $r^*_{ij}$ for $n$ sovereign bonds

\begin{equation}
\overline{r}(t)=\frac{2}{n(n-1)}\sum_{i<j}r^*_{ij}(t),   
\end{equation}

and the variance of correlations $u(t)$ at time-t  

\begin{figure}
	\centering
		{\includegraphics[width=1\textwidth]{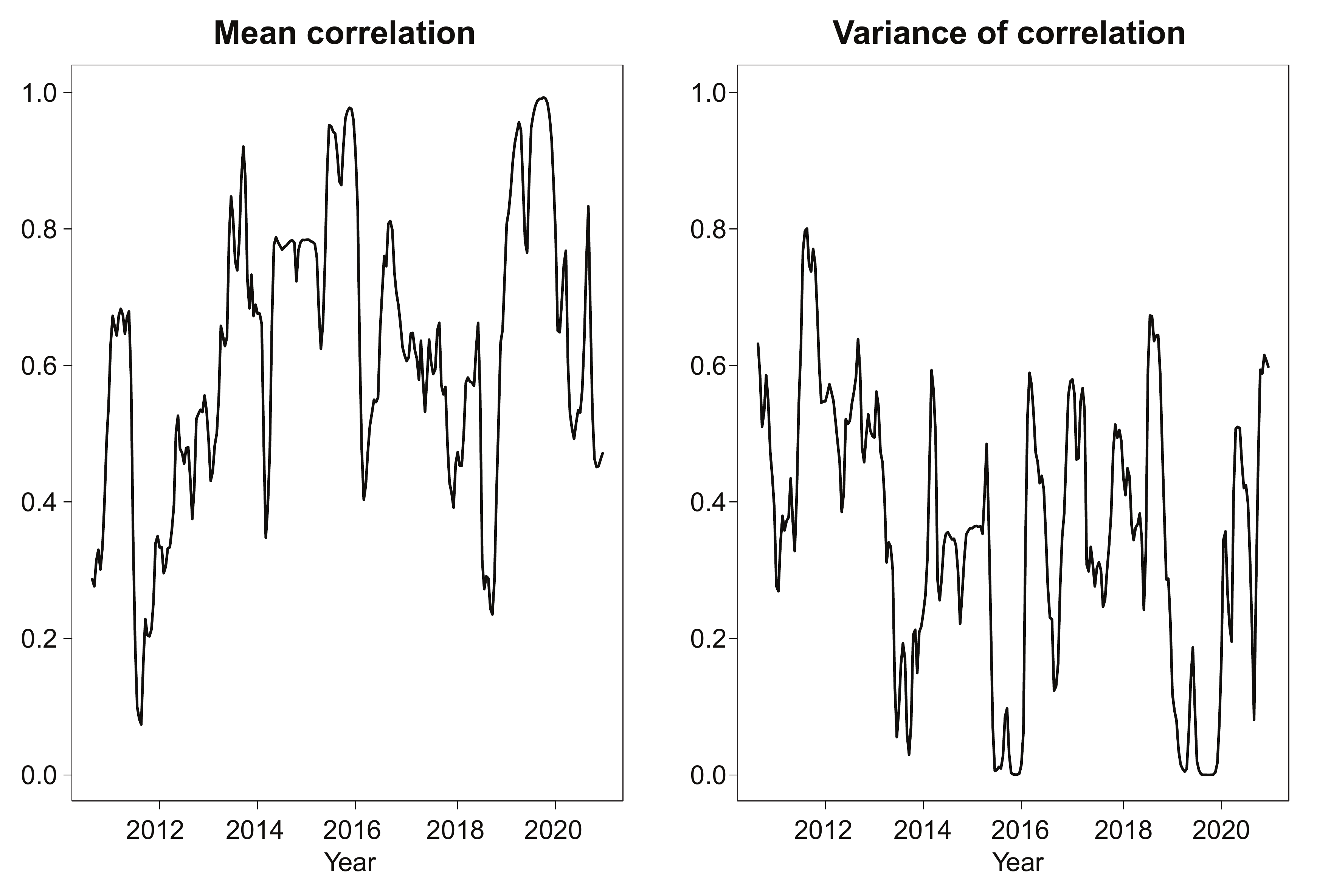}}
\caption{The plots represent the mean and variance of the conditional Pearson correlation matrix. The length of windows is 120 with a displacement value of $\delta=10$ days.}
\label{fig:corrattr}
\end{figure}

\begin{equation}
u(t)=\frac{2}{n(n-1)}\sum_{i<j}(r^*_{ij}(t)-\overline{r}(t))^2. 
\end{equation}

From Figure \ref{fig:corrattr}, we find that the mean correlation $\overline{r}(t)$ is highest at $0.99$ in Oct 2019. This suggests that a COVID-19 impact was a continuation on the decrease of the mean correlation, and throughout the punitive lock down measures introduced by the majority of European countries in Feb-Mar 2020. The decreases in mean correlation are observed within the 2012 period during the European debt crisis, in which several European countries received EU-IMF bailouts to cope with government debt. Within 2016, there was a combination of political uncertainty which followed from the UK and the increased debt accumulation by Italian banks. The variance $u(t)$ also follows a trend similar to the mean correlation, with the smallest variance of $4.48\times10^{-5}$ in October 2019. Within 2020, the variance increases between sovereign bonds and reflects the differences between the correlations of low and high yield.\

\subsection{Network length}

We consider the normalized network length $L(t)$, which is introduced in Onnela, et al~\cite{onnela2003dynamic} as the normalized tree length. We define the measure as the normalized network length, as this measure is considered for AG and TMFG non-tree networks. The network length is a measure of the mean link weights on the subset of links $E(t)$, which are present within the filtered network on the distance matrix at time-t

\begin{equation}
L(t) = \frac{1}{\#\{(i,j)\in E(t)\}}\sum_{(i,j)\in E(t)}d_{ij}(t),
\end{equation}

with the variance $V(t)$ defined on network links\

\begin{figure}
\centering
{\includegraphics[width=1\textwidth]{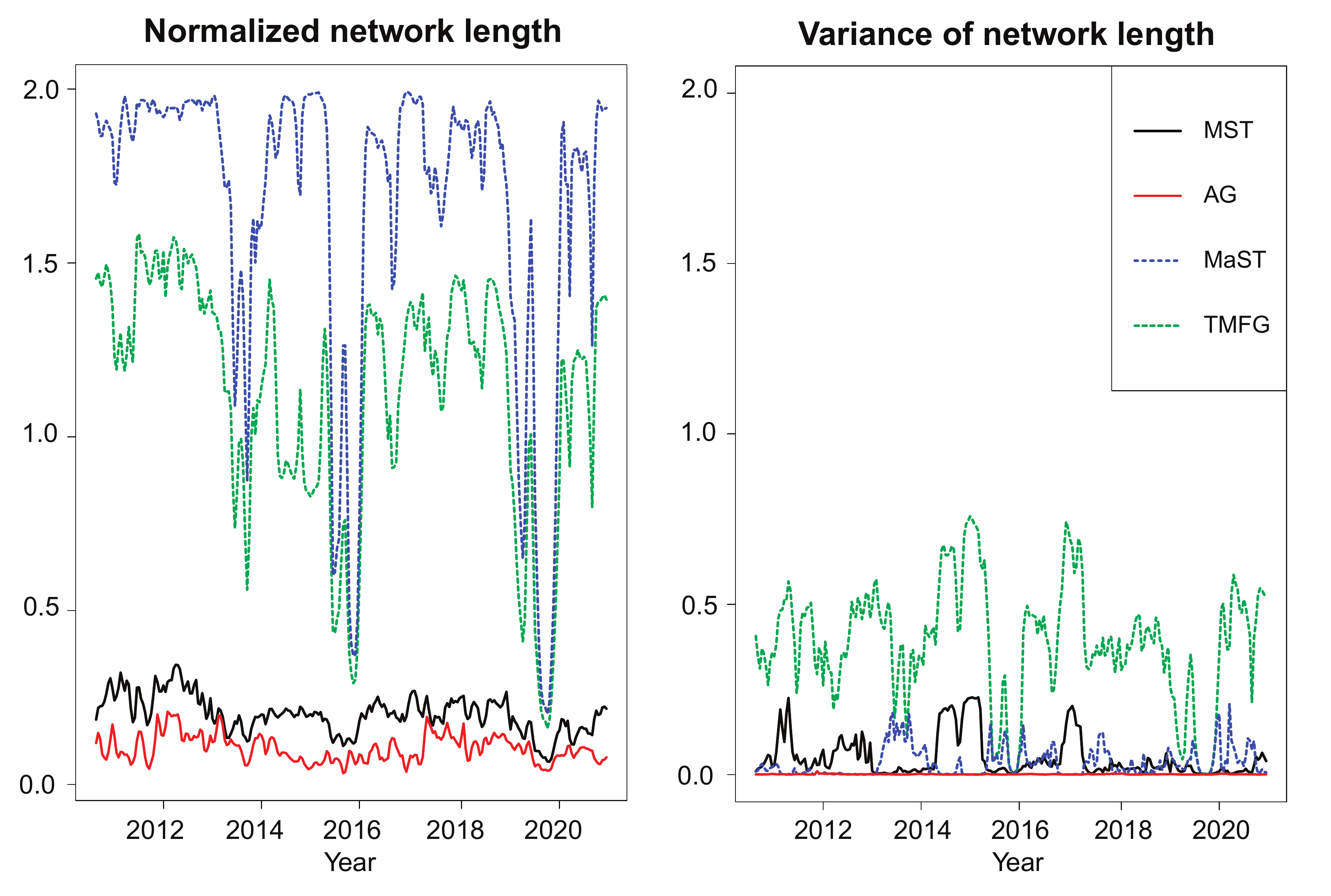}}
\caption{The plots represent the normalized and variance of the network length for MST, TMFG, MaST, and AG networks, with windows of length 120 and $\delta=10$ days.}
\label{fig:treeattr}
\end{figure}

\begin{equation}
V(t) = \frac{1}{\#\{(i,j)\in E(t)\}}\sum_{(i,j)\in E(t)}(d_{ij}(t)-L(t))^2.   
\end{equation}

The plots in Figure \ref{fig:treeattr} represent the mean and variance of the network length. As each filtering method considers a subset of weighted links, the normalized length $L(t)$ is monotonic between all methods and decreases with the increased proportion of positively correlated links within the network. We highlight movements in the normalized network length during the COVID-19 period, which is reflected across all filtering methods. This movement is also observed within 2016,  but only towards a subset of correlations within the MaST and TMFG compared with the MST and AG. The relative difference between the normalized network lengths is least evident in periods of low variance; this is observed in the 2019-2020 period, where the difference between all methods decreases.\ 

We find the variance is highest within the TMFG and lowest with the AG approach. Compared with the mean and variance of the correlation values in Figure \ref{fig:corrattr}, the difference between values within the equivalent network measures is overall higher, particularly within the MaST. There appears to be an overall reciprocal relation between the variance trends of spanning-tree approaches, where both values are small for some periods. When we consider the variance of the AG, the concentration of links, and the adjustment in the conditional correlation results in a flattened trend.    

\begin{figure}
\centering
{\includegraphics[width=1\textwidth]{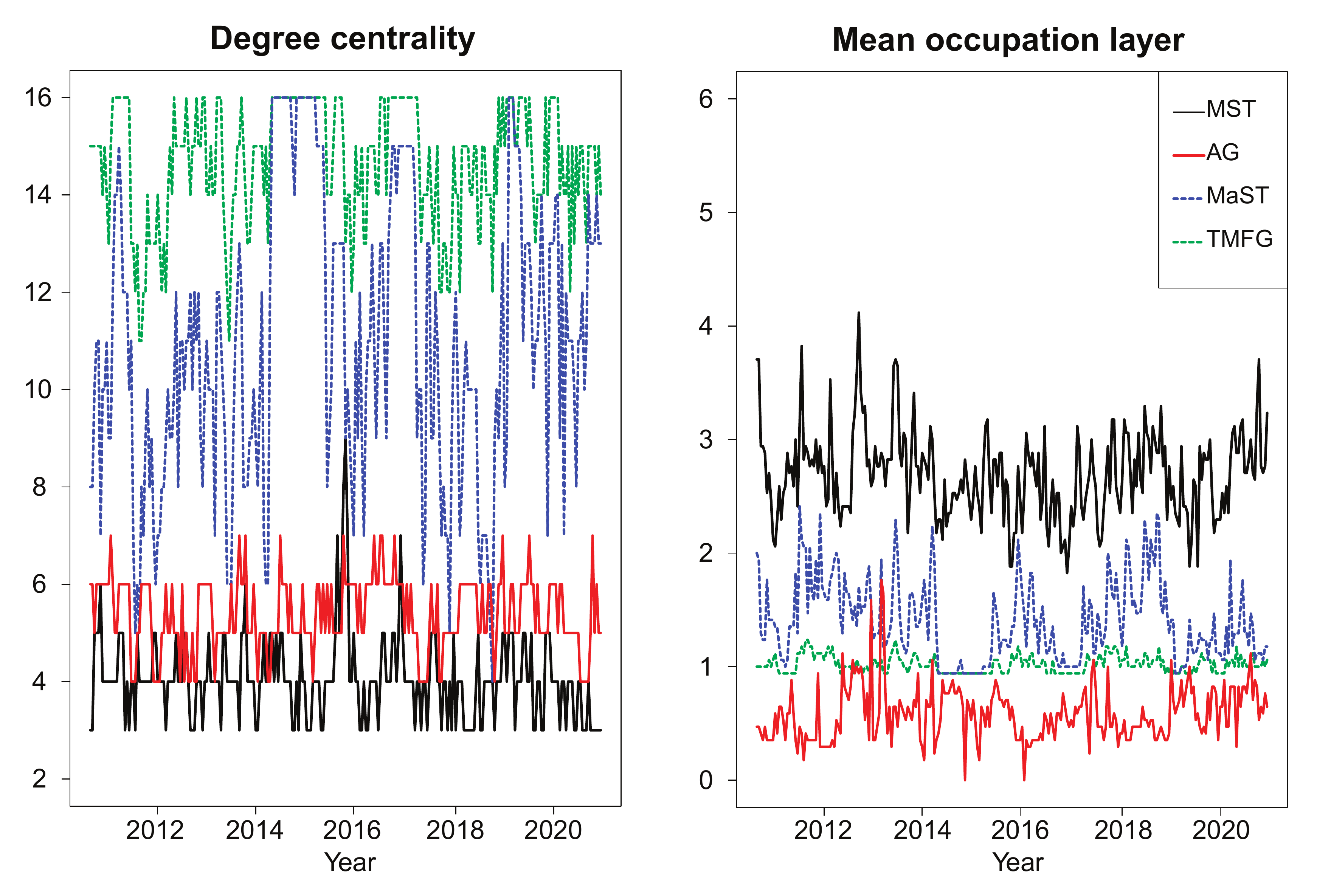}}
\caption{The plots represent the degree centrality and mean occupation layer for MST, TMFG, MaST and AG networks, with windows of length 120 and $\delta=10$ days.}
\label{fig:relsvltree}
\end{figure}

\subsection{Network centrality}

We define the degree centrality for the node of maximum degree $C(t)$ at time-$t$. This measure considers the number of direct links 

\begin{equation}
C(t) = \max_{i\in\{1,...,n\}}\sum_{j\in E(t)}^{n} {1}(d_{ij}>0).
\end{equation}

The mean occupation layer $\eta(t)$ (MOL) introduced in Onnela, et al~\cite{onnela2003dynamic} is a measure of the centrality of the network, relative to the central node $\upsilon(t)$. We define $lev_{i}(t)$ as the level of the node, which is the distance of the node relative to $\upsilon(t)$, where the central node and nodes unconnected relative to the central node have a level value of $0$,

\begin{equation}
  \eta(t) = \frac{1}{n}\sum_{i=1}^{n}lev_{i}(\upsilon(t)).  
\end{equation}

We use the betweenness centrality to define the central node $\upsilon(t)$ for the MOL. Introduced in Freeman~\cite{freeman1977set}, the betweenness $B(t)$ considers the number of shortest paths $s_{ij}(k)$ between $i$ and $j$ which pass through the node $k$, relative to the total number of shortest paths $s_{ij}$ between $i$ and $j$, where $i\neq j \neq k$\ 

\begin{equation}
   B_{k}(t) = \sum_{i\neq k}\sum_{j\neq k, j\neq i}\frac{s_{ij}(k)}{s_{ij}}. 
\end{equation}

Within the MST, the majority of degree centrality ranges between $3$ to $5$, but can be as high as $9$ for some periods. The trend within the MST remains stable, where the central node under degree centrality is associated with multiple sovereign bonds, e.g., Netherlands $11\%$, Portugal $10\%$, and Italy $10\%$ across all periods. The MaST has the highest variation, with a centralized network structure in some periods, e.g., $C(t)$ of 16, forming a star-shaped network structure. This is usually associated with Greece $27\%$, Iceland $25\%$, and Romania $18\%$, which are identified as the central node $70\%$ of the time. The degree centrality on average is naturally highest with the TMFG, under a higher network density, where the central nodes are identified as Iceland and Romania, similar to the MaST. The AG identifies the Netherlands and France within the degree centrality, under a higher proportion of $30\%$ and $13\%$ compared with the MST. Within Figure \ref{fig:relsvltree}, the MOL on average is smallest for the AG, because of the $0$ level values from unconnected nodes, in which an unconnected node is present within all considered windows. We find that all nodes within the TMFG have a maximum path length of $3$ between any two nodes, across all periods. Between the MST and MaST, the MOL is higher within the MaST, where the degree centrality of nodes within the network is higher.\

\section{COVID-19 networks}
\label{sec: plots}

We analyze the temporal changes of sovereign bond yields between Jan 2020 and Dec 2020. This interval establishes a period in which COVID-19 was highly active across multiple European countries. We first construct networks under each of the filtering methods and relate the network topology to economic trends. Then, we implement an exponential random graph model (ERGM) to verify the significance of these explanatory variables within each constructed network. We consider analysis as in \cite{deev2020connectedness}, in which they use an ERGM to analyze the interconnectedness of financial institutions across Europe under different node variables. 

\subsection{Network plots}

\begin{figure}[h!]
\centering
{\includegraphics[width=1\textwidth]{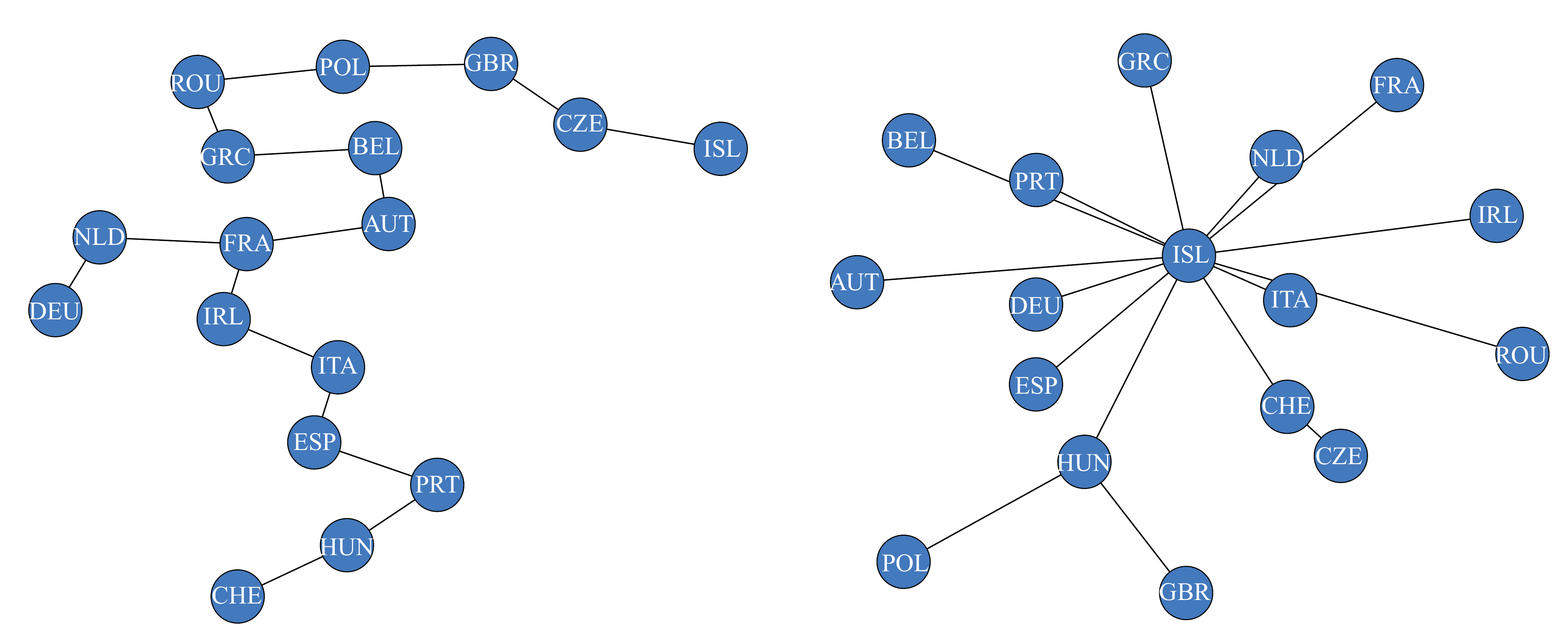}}
\caption{The plots represent the minimum spanning tree (left) and maximum spanning tree (right) for the Jan 2020 - Dec 2020 period.}
\label{fig:SpanPlot}
\end{figure}

Under the MST for the COVID-19 period, we find France has the highest degree centrality of 3. The network also exhibits clusters between a subset of southern European countries, as observed within the connected component of Italy, Portugal and Spain. Within the network, there is a connection between all ABFN countries, but countries within this group also facilitate the connecting component within GIIPS countries, where Belgium is connected with Greek sovereign bonds. The UK and Eastern European countries remain on the periphery, with ABFN countries occupying the core of the network structure. For the MaST in Figure \ref{fig:SpanPlot}, there exists a high degree centrality for Icelandic bond yields. This contrasts to the observed regional hub structure within the MST, where the degree centrality is similar between all nodes. The UK remains within the periphery of the MaST structure when considering anti-correlations, and shows UK bond yields fluctuate less with movements of other European bonds, compared with previous years. This is also observed for sovereign bonds for other countries with non-Euro currencies e.g., Czech Republic and Hungary.\

\begin{figure}
\centering
{\includegraphics[width=1\textwidth]{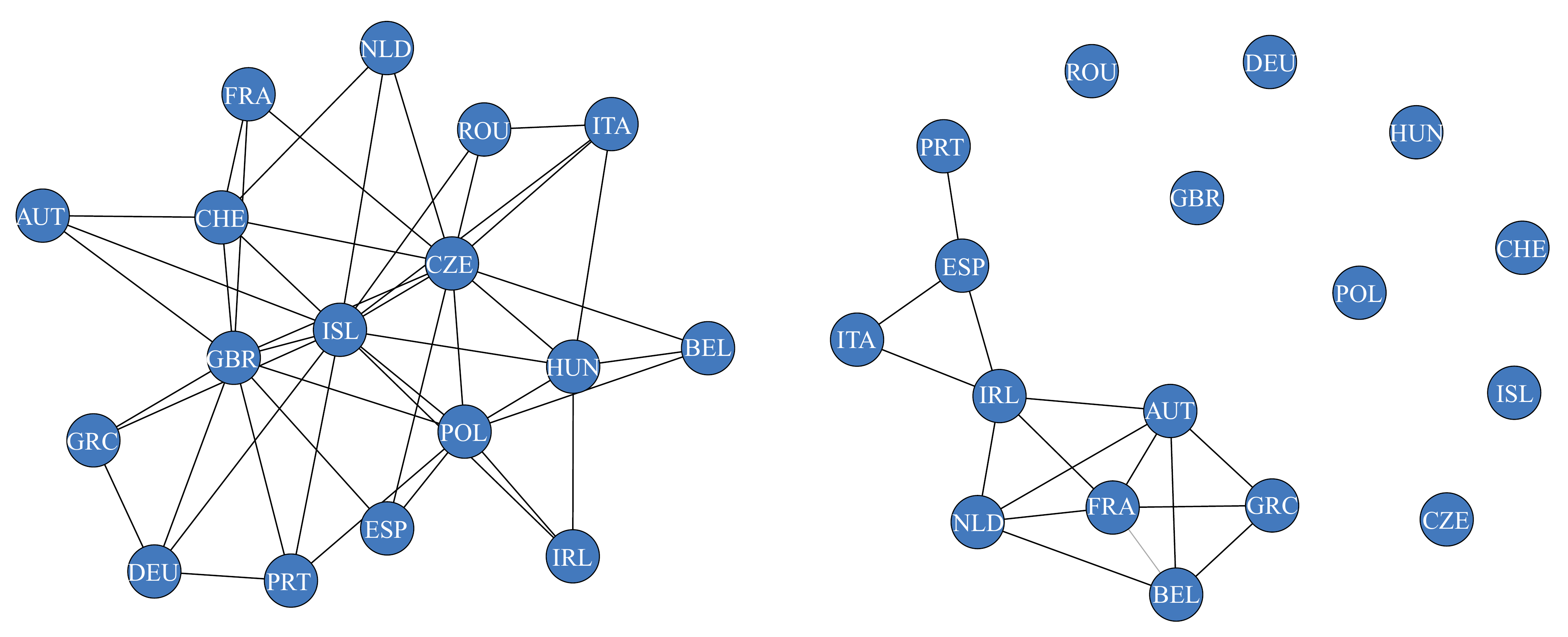}}
\caption{The plots represent the triangulated filtering maximal graph (left) and asset graph (right) for the Jan 2020 - Dec 2020 period.}
\label{fig:GraphPlot}
\end{figure}

We find nodes within the TMFG to have the highest degree in Iceland at $13$ and Czech Republic at $11$. This resembles the links within the MaST, where $75\%$ of links are present within both networks. There is also the associated degree centrality of the MaST, which is observed within the TMFG connected nodes. Under the TMFG, nodes have a higher degree connectivity when considering an increased number of links. This is the case for the UK, which has a degree value of $10$ compared with other filtered networks. We find the AG exhibits one large component which consists of ABFN and GIIPS countries, where the majority of remaining nodes are non-Euro countries and are unconnected within the network. By solely considering the most positive correlations, we include the formation of cliques between countries, which is prevalent within the western European group of 6 nodes. This level of disintegration which is observed during COVID-19 is supported by previous studies of the 2012 Euro zone debt crisis \cite{baur2020decoupling}.\

Under various constraints, there is a commonality between sovereign bonds across network filtering methods. We find for positive correlations, that Euro-zone countries have a high degree centrality, with non-Euro countries are predominately located within the periphery of the network. This is distinctive within the AG, where cliques are only formed between GIIPS and ABFN countries. The anti-correlations within the MaST inform the trends between non-Euro countries and the remaining Euro-area countries. This structure is supported within the TMFG, with the planar graph presenting similarities with the MaST on the degree centrality of nodes i.e., particularly for Iceland.\

\subsection{Exponential Random Graph Model for COVID-19 networks}

We analyze the filtered networks as in Section \ref{sec: plots} under an ERGM. In this approach, we consider a family of random unweighted networks $W$ on the observed network $w$. We define the function of network statistics $z(w)$ and can computationally use the Maximum Likelihood Estimator (MLE) to consider this space of networks. We define the general model for $p$ number of parameters with coefficient values $\theta$ as follows:
\begin{align}
    &\mathbb{P}(W=w)=\frac{\exp\left\{\theta^{T}z(w)\right\}}{\kappa(\theta)},\\
    &\log\left(\exp\{\theta^Tz(w)\}\right)=\theta_{1}z_{1}(w)+\theta_{2}z_{2}(w)+\ldots+\theta_{p}z_{p}(w)
\end{align}

and $\kappa(\theta)$ as the normalizing constant. Although these computations can be expensive for a large number of nodes $n$, we can address these issues by using Markov Chain Monte Carlo (MCMC) methods. We analyze the local interactions between nodes and generate $10,000$ random networks for each filtering method.
    
For the ERGM, we define the "intercept" of the model as the number of links observed within each filtered network. We then consider node level variables under a discrete classification for different economic groups. If the node is within the economic group, we assign a value of "$1$", otherwise the node has a "$0$" value. We also incorporate continuous variables which is represented by economic and health data for all nodes. The data is provided by the International Monetary Fund (IMF) \cite{IMF} and the European Centre for Disease Prevention and Control (ECDC) \cite{ECDC} within the 2020 year. As we consider the coefficient values for different parameters, we further discuss the global model adequacy and fit under the ERGM. We consider the Akaike information criterion (AIC) and Bayesian information criterion (BIC) for the model adequacy, which are measures of the goodness of fit compared with the number of parameters used. We then use the log-likelihood value $LL(\theta)$, and consider the relative difference between the log-likelihood of the ERGM and null model as a measure of the model fit.\ 

\begin{table}[ht]
\centering
\begin{tabular}{cccccccc}
  \hline
 \multicolumn{1}{c}{\textbf{Country}} &
  \multicolumn{1}{c}{\textbf{GIIPS}} &
   \multicolumn{1}{c}{\textbf{ABFN}} &
 \multicolumn{1}{c}{\textbf{Euro}} & 
 \multicolumn{1}{c}{\textbf{COVID-19}}&
\multicolumn{1}{c}{\textbf{Debt}} & 
\multicolumn{1}{c}{\textbf{Inflation}}& 
\multicolumn{1}{c}{\textbf{Account}} \\
 & & & & 
 \multicolumn{1}{c}{\textbf{Deaths (\%)}}&
\multicolumn{1}{c}{\textbf{to GDP}} & 
\multicolumn{1}{c}{\textbf{Rate}}& 
\multicolumn{1}{c}{\textbf{Balance}} \\
  \hline
Austria &  & \checkmark & \checkmark & 0.07 & 84.30 & 1.80 & 2.50 \\
  Belgium &  & \checkmark & \checkmark & 0.17  & 117.10 & 1.20 & -0.80 \\
  Czech &  &  &  & 0.11 & 41.40 & 2.40 & -0.50 \\
  France &  & \checkmark & \checkmark & 0.10  & 118.60 & 0.60 & -1.80 \\ 
  Germany &  &  & \checkmark & 0.04  & 72.20 & 1.10 & 6.80 \\
  Greece & \checkmark &  & \checkmark & 0.05  & 200.50 & 0.70 & -4.50 \\ 
  Hungary &  &  &  & 0.10  & 75.90 & 3.40 & -0.90 \\ 
  Iceland &  &  &  & 0.01  & 52.50 & 2.80 & 0.20 \\
  Ireland & \checkmark &  & \checkmark & 0.05  & 61.30 & 0.60 & 5.50 \\
  Italy & \checkmark &  & \checkmark & 0.12  & 158.30 & 0.60 & 3.00 \\
  Netherlands &  & \checkmark & \checkmark & 0.07  & 61.10 & 1.50 & 9.00 \\ 
  Poland &  &  &  & 0.08  & 60.20 & 2.30 & 1.80 \\ 
  Portugal & \checkmark &  & \checkmark & 0.07  & 130.00 & 1.10 & -3.50 \\
  Romania &  &  &  & 0.08  & 49.60 & 2.50 & -4.50 \\
  Spain & \checkmark &  & \checkmark & 0.11  & 121.30 & 0.80 & 0.90 \\ 
  Switzerland &  &  &  & 0.08  & 48.50 & 0.00 & 9.00 \\ 
  UK &  &  &  & 0.11  & 111.50 & 1.20 & -3.80 \\
   \hline
\end{tabular}
\caption{The following table represents health and economic attributes for countries within the ERGM. The debt is defined as the gross amount relative to the GDP, the inflation rate is recorded for the average consumer prices and the current account balance is the volume of recorded transactions relative to the countries GDP within 2020. We consider the total number of COVID-19 deaths which occurred in 2020 relative to the population size of each country.}
\label{tab:economics}
\end{table}

Within Table \ref{tab:economics}, we define two economic groups of GIIPS and ABFN countries. All of these countries have adopted the Euro and represent the core of connected nodes within the different filtering methods. We use the country debt relative to the GDP, inflation rate, and account balance for 2020 as known economic indicators within the ERGM. As a health indicator of COVID-19, we consider the total number of COVID-19 deaths relative to the size of the country population recorded within the 2020 year. If we compare the relative number of COVID-19 deaths with the debt of countries within Table \ref{tab:economics}, we find several countries have both high levels of COVID-19 deaths and debt to GDP, e.g., Belgium, Italy, and the UK. For other economic indicators, we find an overall negative relationship between inflation and debt, however, some countries, i.e., both Switzerland and Ireland have low inflation and debt value. The account balance of countries is highest with the Netherlands and Switzerland, with Greece and Romania having the most negative values.\

\begin{table}[h!]
\centering
\begin{tabular}{clcccc}
  \hline
  \multicolumn{1}{c}{\textbf{Panel A}} & 
  \multicolumn{1}{l}{\textbf{Parameters}} &
 \multicolumn{1}{c}{\textbf{MST}} &
  \multicolumn{1}{c}{\textbf{TMFG}} &
   \multicolumn{1}{c}{\textbf{MaST}} &
 \multicolumn{1}{c}{\textbf{AG}} \\
\hline
&Edges           & -2.36  & 0.82      & 0.43         & -36.14    \\
&GIIPS           & 0.60   & -0.23     & 1.22         & 20.59     \\
&ABFN            & 0.70   & -0.19     & 1.76         & 22.17     \\
&Euro            & -0.27  & -1.75$^{*}$ & -3.50$^{*}$    & -0.62     \\
&COVID-19 Deaths & 1.77   & -4.93     & -39.20$^{***}$ & -6.95     \\
&Debt to GDP     & -0.00  & 0.01      & 0.02         & -0.01     \\
&Inflation       & 0.01   & 0.12      & 0.71         & -1.14     \\
&Account Balance & -0.04  & 0.02      & 0.11         & 0.00      \\
\hline
\multicolumn{1}{c}{\textbf{Panel B}} & \multicolumn{1}{l}{\textbf{Diagnostics}} & & & &\\
\hline 
 & Goodness of Fit Test: AIC  & 112.92 & 155.20    & 69.38       & 58.41    \\
 & Goodness of Fit Test: BIC  & 136.22 & 178.50    & 92.69       & 81.71    \\
 &Log Likelihood (LL$^M(\theta)$)  & -48.46 & -69.60    & -26.69       & -21.21    \\
 &Log Likelihood (LL$^0(\theta)$)  & -49.26 & -86.33   & -49.26 & -49.26    \\
 &Model Fit  & 1.62\% & 19.38\%    & 45.82\%       & 56.94\%    \\
\hline
\multicolumn{5}{l}{\scriptsize{$^{***}p<0.001$; $^{**}p<0.01$; $^{*}p<0.05$}}
\end{tabular}
\caption{The table represents filtered networks under the ERGM. The goodness of fit is defined under the AIC and BIC, a smaller value represents a higher model adequacy. We define the model fit as $100\times\left[1-\left(LL^M(\theta)/LL^0(\theta)\right)\right]$, where $LL^M(\theta)$ is the log-likelihood of ERGM and $LL^0(\theta)$ is the log-likelihood of the null model. The $LL^M(\theta)$ includes  the link and node level parameters, where the null model $LL^0(\theta)$ only includes the link parameter.}
\label{tab:coefficients}
\end{table}

We find the AG to have the highest model fit under the ERGM and the MST with the lowest fit (see Table \ref{tab:coefficients}). We observe that the GIIPS and ABFN coefficient values are high within the AG, which mainly describe the large component of Euro countries. Under the MST, nodes within the Euro establish links with other Euro countries because of positive correlations and non-Euro countries to satisfy the topological constraints. As the interpreted co-movement is concentrated within Euro countries, the MST removes the representative cliques between nodes, which decreases the coefficient values of GIIPS and ABFN.\

The coefficient of COVID-19 deaths is highly significant within the MaST. Because of the centralized structure around Iceland in which COVID-19 deaths for Iceland are low compared with all other nodes (the death rate of the next lowest value is four times higher than Iceland). As there is a centralized structure when considering negative correlations, we find compatibility of the topological requirement with the ERGM, which is not observed under positives correlations. When the density of the network increases within the TMFG, the network structure decreases the model fit and the model adequacy of the ERGM. We still observe a coefficient value where there is a formation of links between non-Euro countries, i.e., Czech Republic and Iceland with Euro area countries.\

If we consider economic indicators, there is a smaller coefficient value across all parameters compared with COVID-19 deaths. The inflation rate under the MaST is positive between two nodes and represents the links with other countries with high inflation rates. This contrasts with the coefficient value within the AG, which has a negative coefficient between locally connected nodes. Overall, we find altering any one of the conditions within the MST increases the model fitness to the ERGM. This also results in a decrease of the model adequacy in some networks compared with the MST. For negative correlations, we find higher compatibility between the topological structure and model fit under the MaST compared with the MST for positive correlations. Through these approaches, we can capture the core interactions observed between Euro countries and their links with non-Euro countries. We can also factor in economic and health node variables, in which we find COVID-19 deaths to be highly significant.\

\section{Conclusion}
\label{sec: con}

As a response to the COVID-19 pandemic, most countries implemented various socio-economic policies and business restrictions almost simultaneously. An immediate consequence was an increase in yield rates for these nations. The resulting upward co-movement and upward movements in other yield rates explain the decrease in the mean correlation in bond dynamics, coinciding with the pandemic outbreak. Thus, understanding the dynamics of financial instruments in the Euro area is relevant to assess the increased economic strain from events seen in the last decade.\

In this paper, we consider the movements of European sovereign bond yields for network filtering methods, where we focus on the COVID-19 period. We find that the impact of COVID-19 decreased the mean correlation, which was reflected within the normalized network length of all filtering methods. The network topology remained consistent with previous years, in which the trends between approaches were distinctive. The degree centrality was highest for GIIPS and ABFN countries when considering positive correlations and non-Euro countries within negatively correlated type networks. We identified the network structures of filtering methods within the COVID-19 period, which showed one large component consisting of GIIPS and ABFN countries for positive correlations. We were able to verify several of these relationships under an ERGM, in which we find COVID-19 deaths to be significant within negatively correlated networks.\ 

However, depending on the terms of each bond, the European bond market reacted positively after central banks (e.g., Bank of England, European Central Bank, Swiss National Bank) increased their financial programs directed to alleviating the financial pressure on markets and to providing financial liquidity to issuers. Namely, the bond purchase programs had aimed to consolidated market recovery and help to displacing investors toward other financial assets. As a result, prices recovered and remain close to the high levels of the 2020 second quarter, but not at the same level before March's stress situation, especially in 10Y bonds. Additionally, if liquidity provided by central banks starts to drop off, the market dynamics could adjust to economic performance and not its financial performance. In other words, the resulting dynamics could explain an increase in mean correlation in bond dynamics coinciding with the economic dynamics after the pandemic and the increment in yield rates.\ 

Although we consider the sovereign bond yields with a 10Y maturity as a benchmark, this research can be extended to sovereign bonds with different maturities (e.g., short term 1Y, 2Y or 5Y, and long term 20Y or 30Y) because these bonds could reveal interesting effects and confirm that sovereign bonds are a good indicator to identify the economic impact of COVID-19. As each sovereign bond has the different yield and volatility trends, we considered using the zero-coupon curve to evaluate the full extent of COVID-19 on sovereign bonds.\

\small{\bibliography{References}}

\begin{thebibliography}{}

\bibitem[Aldasoro and Alves, 2018]{ALDASORO201817}
Aldasoro, I. and Alves, I. (2018).
\newblock Multiplex interbank networks and systemic importance: An application
  to european data.
\newblock {\em Journal of Financial Stability}, 35:17 -- 37.
\newblock Network models, stress testing and other tools for financial
  stability monitoring and macroprudential policy design and implementation.

\bibitem[Alqaryouti et~al., 2019]{10.1007/978-3-319-99010-1_53}
Alqaryouti, O., Farouk, T., and Siyam, N. (2019).
\newblock Clustering stock markets for balanced portfolio construction.
\newblock In Hassanien, A.~E., Tolba, M.~F., Shaalan, K., and Azar, A.~T.,
  editors, {\em Proceedings of the International Conference on Advanced
  Intelligent Systems and Informatics 2018}, pages 577--587, Cham. Springer
  International Publishing.

\bibitem[Alter and Beyer, 2014]{ALTER2014134}
Alter, A. and Beyer, A. (2014).
\newblock The dynamics of spillover effects during the european sovereign debt
  turmoil.
\newblock {\em Journal of Banking \& Finance}, 42:134 -- 153.

\bibitem[Antonakakis and Vergos, 2013]{antonakakis2013sovereign}
Antonakakis, N. and Vergos, K. (2013).
\newblock Sovereign bond yield spillovers in the euro zone during the financial
  and debt crisis.
\newblock {\em Journal of International Financial Markets, Institutions and
  Money}, 26:258--272.

\bibitem[Aste et~al., 2010]{aste2010correlation}
Aste, T., Shaw, W., and Di~Matteo, T. (2010).
\newblock Correlation structure and dynamics in volatile markets.
\newblock {\em New Journal of Physics}, 12(8):085009.

\bibitem[Balli, 2009]{balli2009spillover}
Balli, F. (2009).
\newblock Spillover effects on government bond yields in euro zone. does full
  financial integration exist in european government bond markets?
\newblock {\em Journal of Economics and Finance}, 33(4):331.

\bibitem[Bargigli et~al., 2016]{Bargigli2016}
Bargigli, L., di~Iasio, G., Infante, L., Lillo, F., and Pierobon, F. (2016).
\newblock {\em Interbank Markets and Multiplex Networks: Centrality Measures
  and Statistical Null Models}, chapter~11, pages 179--194.
\newblock Springer International Publishing, Cham.

\bibitem[Bartolozzi et~al., 2007]{Bartolozzi2007}
Bartolozzi, M., Mellen, C., Di~Matteo, T., and Aste, T. (2007).
\newblock Multi-scale correlations in different futures markets.
\newblock {\em The European Physical Journal B}, 58(2):207--220.

\bibitem[Baur, 2020]{baur2020decoupling}
Baur, D.~G. (2020).
\newblock Decoupling and contagion: The special case of the eurozone sovereign
  debt crisis.
\newblock {\em International Review of Finance}, 20(1):133--154.

\bibitem[Beck et~al., 2016]{BECK2016449}
Beck, R., Georgiadis, G., and Gräb, J. (2016).
\newblock The geography of the great rebalancing in euro area bond markets
  during the sovereign debt crisis.
\newblock {\em Journal of Empirical Finance}, 38:449 -- 460.

\bibitem[Birch et~al., 2016]{birch2016analysis}
Birch, J., Pantelous, A.~A., and Soram{\"a}ki, K. (2016).
\newblock Analysis of correlation based networks representing dax 30 stock
  price returns.
\newblock {\em Computational Economics}, 47(4):501--525.

\bibitem[Claeys and Va{\v{s}}{\'\i}{\v{c}}ek, 2014]{claeys2014measuring}
Claeys, P. and Va{\v{s}}{\'\i}{\v{c}}ek, B. (2014).
\newblock Measuring bilateral spillover and testing contagion on sovereign bond
  markets in europe.
\newblock {\em Journal of Banking \& Finance}, 46:151--165.

\bibitem[Darbellay and Wuertz, 2000]{darbellay2000entropy}
Darbellay, G.~A. and Wuertz, D. (2000).
\newblock The entropy as a tool for analysing statistical dependences in
  financial time series.
\newblock {\em Physica A: Statistical Mechanics and its Applications},
  287(3-4):429--439.

\bibitem[Deev and Ly{\'o}csa, 2020]{deev2020connectedness}
Deev, O. and Ly{\'o}csa, {\v{S}}. (2020).
\newblock Connectedness of financial institutions in europe: A network approach
  across quantiles.
\newblock {\em Physica A: Statistical Mechanics and its Applications}, page
  124035.

\bibitem[Dias, 2012]{dias2012sovereign}
Dias, J. (2012).
\newblock Sovereign debt crisis in the european union: A minimum spanning tree
  approach.
\newblock {\em Physica A: Statistical Mechanics and its Applications},
  391(5):2046--2055.

\bibitem[Dias, 2013]{dias2013spanning}
Dias, J. (2013).
\newblock Spanning trees and the eurozone crisis.
\newblock {\em Physica A: Statistical Mechanics and its Applications},
  392(23):5974--5984.

\bibitem[ecdc.europa.eu, 2020]{ECDC}
ecdc.europa.eu (2020).
\newblock Covid-19 death statistics.

\bibitem[Forbes and Rigobon, 2002]{forbes2002no}
Forbes, K.~J. and Rigobon, R. (2002).
\newblock No contagion, only interdependence: measuring stock market
  comovements.
\newblock {\em The journal of Finance}, 57(5):2223--2261.

\bibitem[Freeman, 1977]{freeman1977set}
Freeman, L.~C. (1977).
\newblock A set of measures of centrality based on betweenness.
\newblock {\em Sociometry}, pages 35--41.

\bibitem[Gilmore et~al., 2010]{gilmore2010}
Gilmore, C.~G., Lucey, B.~M., and Boscia, M.~W. (2010).
\newblock Comovements in government bond markets: A minimum spanning tree
  analysis.
\newblock {\em Physica A: Statistical Mechanics and its Applications},
  389:4875--4886.

\bibitem[Guleva et~al., 2015]{GULEVA2015257}
Guleva, V.~Y., Skvorcova, M.~V., and Boukhanovsky, A.~V. (2015).
\newblock Using multiplex networks for banking systems dynamics modelling.
\newblock {\em Procedia Computer Science}, 66:257 -- 266.
\newblock 4th International Young Scientist Conference on Computational
  Science.

\bibitem[Heimo et~al., 2009]{heimo2009maximal}
Heimo, T., Kaski, K., and Saram{\"a}ki, J. (2009).
\newblock Maximal spanning trees, asset graphs and random matrix denoising in
  the analysis of dynamics of financial networks.
\newblock {\em Physica A: Statistical Mechanics and its Applications},
  388(2-3):145--156.

\bibitem[Huang et~al., 2012]{huang2012multifractal}
Huang, J., Shang, P., and Zhao, X. (2012).
\newblock Multifractal diffusion entropy analysis on stock volatility in
  financial markets.
\newblock {\em Physica A: Statistical Mechanics and its Applications},
  391(22):5739--5745.

\bibitem[IMF.org, 2020]{IMF}
IMF.org (2020).
\newblock Imf world economic outlook.

\bibitem[Investing.com, 2020]{investing}
Investing.com (2020).
\newblock World government bonds.

\bibitem[Isogai, 2017]{Isogai2017}
Isogai, T. (2017).
\newblock Dynamic correlation network analysis of financial asset returns with
  network clustering.
\newblock {\em Applied Network Science}, 2(1):8.

\bibitem[Jang et~al., 2011]{jang2011currency}
Jang, W., Lee, J., and Chang, W. (2011).
\newblock Currency crises and the evolution of foreign exchange market:
  Evidence from minimum spanning tree.
\newblock {\em Physica A: Statistical Mechanics and its Applications},
  390(4):707--718.

\bibitem[Junior et~al., 2015]{Junior_2015}
Junior, L., Mullokandov, A., and Kenett, D. (2015).
\newblock Dependency relations among international stock market indices.
\newblock {\em Journal of Risk and Financial Management}, 8(2):227–265.

\bibitem[Junior and Franca, 2012]{junior2012correlation}
Junior, L.~S. and Franca, I. D.~P. (2012).
\newblock Correlation of financial markets in times of crisis.
\newblock {\em Physica A: Statistical Mechanics and its Applications},
  391(1-2):187--208.

\bibitem[Kok and Montagna, 2016]{RePEc:ecb:ecbwps:20161944}
Kok, C. and Montagna, M. (2016).
\newblock {Multi-layered interbank model for assessing systemic risk}.
\newblock Working Paper Series 1944, European Central Bank.

\bibitem[Kruskal, 1956]{kruskal1956shortest}
Kruskal, J.~B. (1956).
\newblock On the shortest spanning subtree of a graph and the traveling
  salesman problem.
\newblock {\em Proceedings of the American Mathematical Society}, 7(1):48--50.

\bibitem[Kukreti et~al., 2020]{kukreti2020perspective}
Kukreti, V., Pharasi, H.~K., Gupta, P., and Kumar, S. (2020).
\newblock A perspective on correlation-based financial networks and entropy
  measures.
\newblock {\em arXiv preprint arXiv:2004.09448}.

\bibitem[Laloux et~al., 2000]{laloux2000random}
Laloux, L., Cizeau, P., Potters, M., and Bouchaud, J.-P. (2000).
\newblock Random matrix theory and financial correlations.
\newblock {\em International Journal of Theoretical and Applied Finance},
  3(03):391--397.

\bibitem[Leon et~al., 2014]{leon2014}
Leon, C., Leiton, K., and Perez, J. (2014).
\newblock Extracting the sovereigns’ cds market hierarchy: A
  correlation-filtering approach.
\newblock {\em Physica A: Statistical Mechanics and its Applications},
  415:407--420.

\bibitem[Li et~al., 2019]{Li2019}
Li, Y., Jiang, X.-F., Tian, Y., Li, S.-P., and Zheng, B. (2019).
\newblock Portfolio optimization based on network topology.
\newblock {\em Physica A: Statistical Mechanics and its Applications},
  515:671--681.

\bibitem[Maeng et~al., 2012]{maeng2012}
Maeng, S.~E., Choi, H.~W., and Lee, J.~W. (2012).
\newblock Complex networks and minimal spanning trees in international trade
  network.
\newblock {\em International Journal of Modern Physics: Conference Series},
  16:50--60.

\bibitem[Mantegna, 1999]{mantegna1999hierarchical}
Mantegna, R.~N. (1999).
\newblock Hierarchical structure in financial markets.
\newblock {\em The European Physical Journal B-Condensed Matter and Complex
  Systems}, 11(1):193--197.

\bibitem[Massara et~al., 2016]{massara2016network}
Massara, G.~P., Di~Matteo, T., and Aste, T. (2016).
\newblock Network filtering for big data: Triangulated maximally filtered
  graph.
\newblock {\em Journal of Complex Networks}, 5(2):161--178.

\bibitem[Matteo et~al., 2005]{DIMATTEO200521}
Matteo, T.~D., Aste, T., Hyde, S., and Ramsden, S. (2005).
\newblock Interest rates hierarchical structure.
\newblock {\em Physica A: Statistical Mechanics and its Applications},
  355(1):21 -- 33.
\newblock Market Dynamics and Quantitative Economics.

\bibitem[Musmeci et~al., 2015]{10.1371/journal.pone.0116201}
Musmeci, N., Aste, T., and Di~Matteo, T. (2015).
\newblock Relation between financial market structure and the real economy:
  Comparison between clustering methods.
\newblock {\em PLOS ONE}, 10(3):1--24.

\bibitem[Musmeci et~al., 2017]{musmeci2017multiplex}
Musmeci, N., Nicosia, V., Aste, T., Di~Matteo, T., and Latora, V. (2017).
\newblock The multiplex dependency structure of financial markets.
\newblock {\em Complexity}, 2017.

\bibitem[Onnela et~al., 2003a]{onnela2003dynamic}
Onnela, J.-P., Chakraborti, A., Kaski, K., and Kertesz, J. (2003a).
\newblock Dynamic asset trees and black monday.
\newblock {\em Physica A: Statistical Mechanics and its Applications},
  324(1-2):247--252.

\bibitem[Onnela et~al., 2003b]{onnela2003asset}
Onnela, J.-P., Chakraborti, A., Kaski, K., Kertesz, J., and Kanto, A. (2003b).
\newblock Asset trees and asset graphs in financial markets.
\newblock {\em Physica Scripta}, 2003(T106):48.

\bibitem[Onnela et~al., 2004]{onnela2004clustering}
Onnela, J.-P., Kaski, K., and Kert{\'e}sz, J. (2004).
\newblock Clustering and information in correlation based financial networks.
\newblock {\em The European Physical Journal B}, 38(2):353--362.

\bibitem[Plerou et~al., 2002]{plerou2002random}
Plerou, V., Gopikrishnan, P., Rosenow, B., Amaral, L. A.~N., Guhr, T., and
  Stanley, H.~E. (2002).
\newblock Random matrix approach to cross correlations in financial data.
\newblock {\em Physical Review E}, 65(6):066126.

\bibitem[Poledna et~al., 2015]{POLEDNA201570}
Poledna, S., Molina-Borboa, J.~L., Martínez-Jaramillo, S., [van~der Leij], M.,
  and Thurner, S. (2015).
\newblock The multi-layer network nature of systemic risk and its implications
  for the costs of financial crises.
\newblock {\em Journal of Financial Stability}, 20:70 -- 81.

\bibitem[Qian et~al., 2010]{qian2010universal}
Qian, M.-C., Jiang, Z.-Q., and Zhou, W.-X. (2010).
\newblock Universal and nonuniversal allometric scaling behaviors in the
  visibility graphs of world stock market indices.
\newblock {\em Journal of Physics A: Mathematical and Theoretical},
  43(33):335002.

\bibitem[Sandoval~Jr, 2012]{sandoval2012pruning}
Sandoval~Jr, L. (2012).
\newblock Pruning a minimum spanning tree.
\newblock {\em Physica A: Statistical Mechanics and its Applications},
  391(8):2678--2711.

\bibitem[Schwendner et~al., 2015]{Schwendneretal2015}
Schwendner, P., Schuele, M., Ott, T., and Hillebrand, M. (2015).
\newblock European government bond dynamics and stability policies: taming
  contagion risks.
\newblock {\em Journal of Network Theory in Finance}, 1(4):1--25.

\bibitem[Situngkir and Surya, 2005]{situngkir2005stock}
Situngkir, H. and Surya, Y. (2005).
\newblock On stock market dynamics through ultrametricity of minimum spanning
  tree.
\newblock Technical report, Bandung Fe Institute.

\bibitem[Stavroglou et~al., 2016]{Stavroglou2016}
Stavroglou, S.~K., Pantelous, A.~A., Soramaki, K., and Zuev, K. (2016).
\newblock Causality networks of financial assets.
\newblock {\em Journal of Network Theory in Finance}, 3:17--67.

\bibitem[Tang et~al., 2018]{Tang2018}
Tang, Y., Xiong, J.~J., Jia, Z.-Y., and Zhang, Y.-C. (2018).
\newblock Complexities in financial network topological dynamics: Modeling of
  emerging and developed stock markets.
\newblock {\em Complexity}, 2018:4680140.

\bibitem[Tonzer, 2015]{TONZER201519}
Tonzer, L. (2015).
\newblock Cross-border interbank networks, banking risk and contagion.
\newblock {\em Journal of Financial Stability}, 18:19 -- 32.

\bibitem[Tumminello et~al., 2005]{tumminello2005tool}
Tumminello, M., Aste, T., Di~Matteo, T., and Mantegna, R.~N. (2005).
\newblock A tool for filtering information in complex systems.
\newblock {\em Proceedings of the National Academy of Sciences},
  102(30):10421--10426.

\bibitem[Tumminello et~al., 2010]{TUMMINELLO201040}
Tumminello, M., Lillo, F., and Mantegna, R.~N. (2010).
\newblock Correlation, hierarchies, and networks in financial markets.
\newblock {\em Journal of Economic Behavior and Organization}, 75(1):40 -- 58.

\bibitem[Verma et~al., 2019]{doi:10.1080/14697688.2018.1535183}
Verma, A., Buonocore, R.~J., and Matteo, T.~D. (2019).
\newblock A cluster driven log-volatility factor model: a deepening on the
  source of the volatility clustering.
\newblock {\em Quantitative Finance}, 19(6):981--996.

\bibitem[Vodenska et~al., 2016]{Vodenska_2016}
Vodenska, I., Becker, A., Zhou, D., Kenett, D., Stanley, H., and Havlin, S.
  (2016).
\newblock Community analysis of global financial markets.
\newblock {\em Risks}, 4(2):13.

\bibitem[Wang et~al., 2017]{Wang2017}
Wang, G.-J., Xie, C., and Chen, S. (2017).
\newblock Multiscale correlation networks analysis of the us stock market: a
  wavelet analysis.
\newblock {\em Journal of Economic Interaction and Coordination},
  12(3):561--594.

\bibitem[Wu et~al., 2018]{Wu2018}
Wu, S., He, J., Li, S., and Wang, C. (2018).
\newblock Network formation in a multi-asset artificial stock market.
\newblock {\em The European Physical Journal B}, 91(4):66.

\end{thebibliography}

\end{document}